# Super-resolution Imaging by Evanescent Wave Coupling to Surface States on Effective Gain Media

Prateek Mehrotra[1], Chris A. Mack[2], Richard J. Blaikie[1,*]

[1]The MacDiarmid Institute for Advanced Materials and Nanotechnology,

Department of Electrical and Computer Engineering, University of Canterbury,

Christchurch, New Zealand

[2]Department of Chemical Engineering, The University of Texas at Austin, Austin, TX

*Current Address: Department of Physics, University of Otago, PO Box 56, Dunedin, New Zealand. Corresponding author, email: richard.blaikie@otago.ac.nz

## ABSTRACT

Higher resolution demands for semiconductor lithography may be fulfilled by higher numerical aperture (NA) systems. However, NAs more than the photoresist refractive index (~1.7) cause surface confinement of the image. In this letter we describe how evanescent wave coupling to effective gain medium surface states beneath the imaging layer can counter this problem. At λ=193 nm a layer of sapphire on $SiO_2$ counters image decay by an effective-gain-medium resonance phenomena allowing evanescent interferometric lithography to create high aspect ratio structures at *NA*s of 1.85 (26-nm) and beyond.

Advances in super-resolution optical imaging are being aggressively pursued for applications in imaging, data storage and lithography. In biological imaging, for example, recent advances have seen reports of molecular-scale resolution using a range of nonlinear and stochastic processes [1,2], and commercial instruments are now available. For other applications—such as lithography imaging that is the subject of this paper—analogous techniques are not readily applied and alternative means of achieving super-resolved imaging must be developed.

The semiconductor industry has historically reduced the feature sizes used in integrated circuits by a factor of 2 every 4 to 6 years. By far the most important feature to reduce is the pitch of a repeating pattern of lines and spaces, since this determines the packing density of transistors in the circuit, and thus the cost and capability of the device. The resolution ($R$) is defined as the minimum half-pitch and in optical lithography is limited by the ratio of the wavelength ($\lambda$) to the numerical aperture of the imaging tool ($NA$):

$$R = k_1 \frac{\lambda}{NA}, \quad k_1 \geq 0.25 \tag{1}$$

One way to achieve the lowest possible value for the resolution factor $k_1$ (= 0.25) is to interfere two plane waves traveling at the opposite ends of the full $NA$, an approach known as interferometric lithography (IL) [3]. To further improve resolution, one must lower the wavelength or increase $NA$. Today, a wavelength of 193 nm is common and lowering it has proven difficult in manufacturing. Increasing $NA$ is also difficult, being limited by the lowest refractive index in the optical path on the wafer-side of the imaging tool. Today, water is used as the immersion fluid, giving a maximum practical $NA$ of ~1.35. Higher index immersion and lens materials are presently employed in what is known as a solid-immersion system [4]. The limiting refractive index is then that of the photoresist, limiting $NA$ to ~1.6. Higher $NA$s are extremely attractive and feasible but put the system into the evanescent regime. Unfortunately, evanescent images decay exponentially into the resist, leading to images that are practically useful only for extremely thin resists, making pattern transfer difficult. We report new principles, derivations and designs for a method whereby coupling to surface states on an underlying effective gain medium can be used to significantly enhance the depth of field of evanescent images with transverse electric (TE) polarized light. The novelty here, compared to other surface-state reflection proposals, is that such effective gain media can be implemented using readily available materials and standard thin-film deposition techniques; the solution is a feasible one which has not been considered before.

A primary goal of this work is to derive the limits of resist thickness for practical evanescent imaging and to devise schemes to increase these limits so that sufficient thickness of resist may be used to meet the needs of manufacturing. Keeping this in mind, we will limit our solutions to TE polarized light that the industry prefers to use due to the resulting high contrast of the images at all *NA*s. We consider a two-wave interference lithography setup as shown in Fig. 1 with two plane waves interfering at angles $\pm\theta_1$ relative to the normal direction of a planar layer of photoresist and substrate materials.

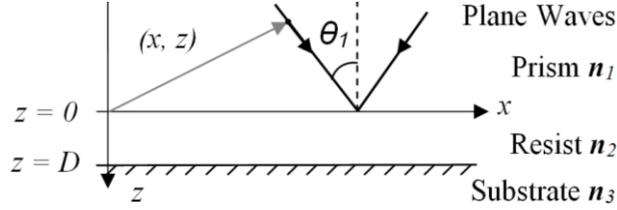

FIG. 1 Interferometric imaging system geometry.

The optical parameters of each medium in the multi-layer imaging systems are more easily dealt with when defined in terms of their complex optical admittance ($\eta$) as opposed to the complex refractive index ($n = n + i\kappa$). For TE polarization the optical admittance of the $i^{th}$ layer in the system is,

$$\eta_i = \sqrt{n_i^2 - (NA)^2} = n_i \cos(\theta_i) \qquad (2)$$

Using this, the Fresnel reflection between layers $i$ and $j$ is given as,

$$\rho_{ij} = \frac{n_i \cos(\theta_i) - n_j \cos(\theta_j)}{n_i \cos(\theta_i) + n_j \cos(\theta_j)} = \frac{\eta_i - \eta_j}{\eta_i + \eta_j} \qquad (3)$$

In the photoresist, the electric-field profile in the $z$ direction as derived in [5] is

$$E_{sw}(\theta,z) = e^{ik_2 z \cos\theta_2} + \rho_{23}(\theta)\tau_D^2(\theta)e^{-ik_2 z \cos\theta_2} \qquad (4)$$

with the exception of a scaling factor, and with $\tau_D = e^{ik_2 D \cos\theta_2}$ and $k_i = 2\pi n_i / \lambda$. The intensity is then simply given as

$$I_{sw}(\theta,z) = E_{sw} E_{sw}^* \qquad (5)$$

Solving for the real and imaginary parts of the resist optical admittance in terms of the real and imaginary parts of the resist refractive index,

$$\text{Re}\{\eta_2\} = \eta_{2R} = \frac{n_2 \kappa_2}{\eta_{2I}} \tag{6}$$

$$\eta_{2I} = \sqrt{NA^2 - n_2^2 + \kappa_2^2} \sqrt{\frac{1}{2}\left[\sqrt{1 + \left(\frac{2n_2\kappa_2}{NA^2 - n_2^2 + \kappa_2^2}\right)^2} + 1\right]} \tag{7}$$

For the weakly absorbing resist approximation ($\kappa_2 \ll n_2$), and assuming $NA$ sufficiently larger than $n_2$, the expression for $\eta_{2I}$ simplifies to

$$\eta_{2I} \approx \sqrt{NA^2 - n_2^2} + \frac{1}{2}\frac{(n_2^2 + 1)\kappa_2^2}{\sqrt{NA^2 - n_2^2}} \approx \sqrt{NA^2 - n_2^2} \tag{8}$$

Using Eq. (6) in Eq. (4),

$$e^{ik_2 z \cos\theta_2} = e^{-2\pi\eta_{2I} z/\lambda} e^{i2\pi n_2 z(\kappa_2/\eta_{2I})/\lambda} \tag{9}$$

Thus, in the evanescent regime, the resist has an effective absorption coefficient given by

$$\alpha_e = \frac{4\pi\eta_{2I}}{\lambda} \tag{10}$$

As an example, consider an evanescent interferometric imaging case where $\lambda = 193$ nm, $NA = 1.85$, and $\mathbf{n_2} = 1.7 + i0.02$. The optical admittance is $\boldsymbol{\eta_2} = 0.047 + i0.73$ and the effective evanescent absorption coefficient is 47.6 µm$^{-1}$. The difference between propagating and evanescent regimes is dramatic. In the evanescent regime, the resist is highly (36 times more) "absorbing", but has a very small real part of the optical admittance. The imaginary part of the optical admittance is governed by how much the $NA$ exceeds the refractive index of the resist. The real part of the optical admittance is proportional to the imaginary part of the resist refractive index and thus can be increased by adding an absorbing dye in the resist. The resulting standing wave intensity now becomes

$$I_{sw}(z) = e^{-\alpha_e z} + |\rho_{23}|^2 e^{-\alpha_e(2D-z)} + 2|\rho_{23}|e^{-\alpha_e D}\cos(4\pi\eta_{2R}(D-z)/\lambda + \phi_{23}) \tag{11}$$

where $D$ is the resist thickness and $\phi_{23}$ is the angle of the substrate reflectivity, given by

$$\rho_{23} = |\rho_{23}|e^{i\phi_{23}} \quad (12)$$

Conventional lithography aims to minimize $\rho_{23}$ to avoid standing waves in resist. In addition, evanescent images at the *NA*s we consider have never been a prime focus in lithography. We propose to set $\rho_{23}$ to an optimum level to counter image decay and develop theory below to show what needs to be done to achieve this.

**SURFACE STATES**

A few articles [6,7,8] have proposed and analyzed how plasmonic effects from a metal underlay may be able to counter evanescent field decay in near-field imaging. An example is presented to set the scene for the new approach we present here. E-field confinement and enhancement through resonant interaction is characteristic of a surface state [9] and Fig. 2 shows such enhancement using plasmonic surface states on a fictitious metal. The exponential intensity decay in a transverse-magnetic (TM) polarized evanescent-wave image (Fig. 2(a)) can be overcome, or even overwhelmed, by resonant coupling to such plasmonic states (Figs. 2(b) and (c)).

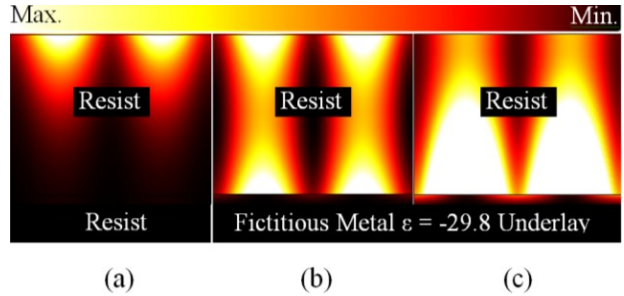

FIG. 2 Evanescent IL with TM polarized light in 82.5-nm thick resist, λ=193 nm with (a) resist underlay, (b) metal (ε = -29.8) underlay, optimal off-resonant enhancement (NA = 1.85), and (c) metal (ε = -29.8) underlay, non-optimal resonant enhancement (NA = 1.79).

The underlying principle is energy extraction which would otherwise totally internally reflect back into the prism and its redistribution in the resist cavity. In fact, the superlens suggested by John Pendry [10] also utilizes this principle and enhancement is achieved via field redistribution by coupling to surface states on both sides of the superlens allowing energy extraction and redistribution from the source apertures to photoresist. The principle is evident by reference to the attenuated total reflectance (ATR) spectrum of Fig. 3(a); setting the operating *NA* around the resonant dip in the ATR spectrum allows fine control of the field profiles to produce a symmetrical intensity distribution (Fig. 2(b)).

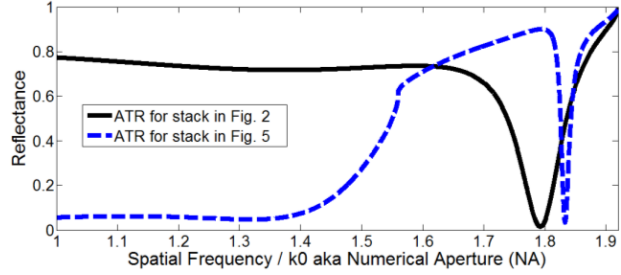

FIG. 3 Attenuated Total Reflectance Spectra for stack in (a) Fig. 2(b,c), (b) Fig. 5 (discussed in a later section).

Here we propose a new approach to such surface-state-enhanced imaging, whereby either TE or TM polarization can be employed and a wider range of real materials can be used, via understanding this concept and analyzing the Fresnel reflection equations in the following sections.

**FRESNEL REFLECTION**

The local enhancement at a metal-resist interface is also evident from the Fresnel reflection equation. Solving the TM case for a reflectivity > 1 results in metals as one solution. However, the enhancement is limited to the TM polarization. Here, we proceed to solve the TE Fresnel reflection for a reflectivity > 1, knowing very well that SPPs, *i.e.* metals, are not a possible solution. We will limit our analysis to non-magnetic media.

$$|\rho_{23}| = \left| \frac{(\eta_{2R} + i\eta_{2I}) - (\eta_{3R} + i\eta_{3I})}{(\eta_{2R} + i\eta_{2I}) + (\eta_{3R} + i\eta_{3I})} \right| > 1 \tag{13}$$

Rearranging real and imaginary terms, taking the absolute squares gives,

$$(\eta_{2R} - \eta_{3R})^2 + (\eta_{2I} - \eta_{3I})^2 > (\eta_{2R} + \eta_{3R})^2 + (\eta_{2I} + \eta_{3I})^2 \tag{14}$$

Cancelling common terms and invoking the assumption of a low-loss resist, we are left with

$$\eta_{2I}\eta_{3I} < 0 \quad or \quad \eta_{3I} < 0 \tag{15}$$

since $\eta_{2I}$ is positive. To satisfy $\eta_{3I} < 0$ we choose the square root in (2) corresponding to $\kappa_3 < 0$. This implies a negative loss or gain in the underlying medium and ensures our solution is restricted to $n_3 > 0$, i.e. non-negative refracting media. While a gain medium below the photoresist may seem at first unfeasible, there is a practical solution that produces reflection > 1 thanks to the evanescent nature of the waves.

**EFFECTIVE GAIN MEDIUM**

Consider a high-index dielectric capable of supporting an ultra-high *NA* in a propagating form. Similar to a waveguide, the fields may be contained within it by satisfying relevant boundary conditions, *i.e.* if the high-index medium is sandwiched between low-index media that cannot support the ultra-high *NA*s. Control over thickness and index of the high index medium allows selection of the spatial frequency that is resonated. Figure 4(b) illustrates this arrangement with approximate field profiles within the media. The ATR spectrum would also show the characteristic dip at resonance, similar to that shown in Fig. 3. In the SPP case, E-fields are confined and enhanced at the interface while decaying into surrounding media [Fig. 4(a)]. Here, a similar purpose is fulfilled by a pseudo-interface [Fig. 4(b)].

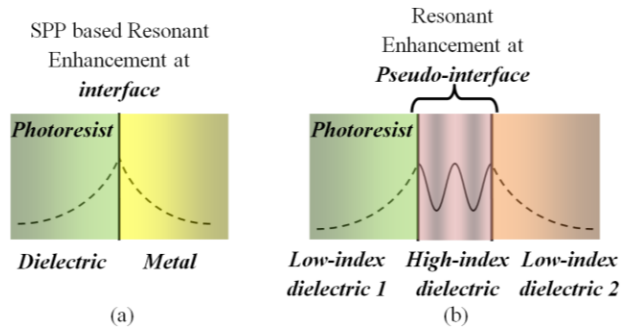

FIG. 4 Evanescent wave enhancement (a) at a metal-dielectric interface through SPP resonance, and (b) at a pseudo-interface formed by sandwiching a high index dielectric between two low-index dielectrics.

Now, relating this back to the result of the previous analysis, Eq. (15), the high index on low index stack is in fact an effective gain medium (EGM) using effective medium theory (EMT) provided we concern ourselves with only reflection at the particular *NA*. The resonant spatial frequency here is the effective surface state. With the TM polarization, such a stack has two EMT solutions: 1) an EGM, and 2) a metal. Knowing that such an arrangement is required, we now continue with the analysis.

**LIMITS - REFLECTIVITY, NUMERICAL APERTURE AND RESIST THICKNESS**

The desired image is that in which the intensity at the bottom of the resist equals the intensity at the top, in order to achieve a symmetrical image profile. For the typical case of small $\eta_{2R}$,

$$I_{sw}(z=0) = 1 + |\rho_{23}|^2 e^{-\alpha_e 2D} + 2|\rho_{23}|e^{-\alpha_e D} \cos(\phi_{23}) \tag{16}$$

$$I_{sw}(z=D) = e^{-\alpha_e D}\left(1 + |\rho_{23}|^2 + 2|\rho_{23}|\cos(\phi_{23})\right)$$

The value of the substrate reflectivity that provides this solution is

$$|\rho_{23}|^2 = \frac{e^{\alpha_e D} - 1}{1 - e^{-\alpha_e D}} \approx e^{\alpha_e D} \tag{17}$$

For this condition, the absorbance limitation will be determined by the intensity of light at the middle of the resist. The goal will be to keep this ratio as close to 1 as possible:

$$\frac{I_{sw}(z=0) = I_{sw}(z=D)}{I_{sw}(z=D/2)} = \frac{1 + |\rho_{23}|^2 + 2|\rho_{23}|\cos(\phi_{23})}{e^{\alpha_e D/2} + |\rho_{23}|^2 e^{-\alpha_e D/2} + 2|\rho_{23}|\cos(\phi_{23})} \tag{18}$$

At the optimum substrate reflectivity,

$$\frac{I_{sw}(z=0)}{I_{sw}(z=D/2)} = \frac{\frac{1}{2}\left(e^{-\alpha_e D/2} + e^{\alpha_e D/2}\right) + \cos(\phi_{23})}{1 + \cos(\phi_{23})} \tag{19}$$

If this ratio of intensities is allowed to reach 2.5 (a reasonable value based on experience), then $(\alpha_e D)_{max} = 4.1$, using the best case $\phi_{23} = 0$. This allows the resist thickness to be twice as great as could be achieved with an ideal conventional mirror $(|\rho_{23}|=1)$. Remarkably, to achieve any possible NA, the constraint on the minimum aspect ratio (AR) (feature height/width) becomes

$$AR_{min} \leq \frac{(\alpha_e D)_{max}}{\pi} = 1.3 \tag{20}$$

While not as high as desired, an AR of 1.3 is reasonable. If the minimum AR is set to 2.0, then the highest NA achievable is $1.32 n_2$, or about 2.2 for a typical photoresist.

**DESIGN & IMPLEMENTATION**

Consider now building the stack proposed in Fig. 4, using a thin film (layer 3) placed between the resist and a semi-infinite substrate (layer 4). The reflectivity between the resist and the film stack, $\rho_{23}'$, will be

$$\rho_{23}' = \frac{\rho_{23} + \rho_{34} \tau_{D3}^2}{1 + \rho_{23} \rho_{34} \tau_{D3}^2} \tag{21}$$

An infinite reflectivity can be obtained if $\rho_{23} \rho_{34} \tau_{D3}^2 = -1$. This requires first that $\left|\tau_{D3}^2\right| = 1$, so that layer 3 must be non-absorbing and must be operating in a propagating (not evanescent) regime. Thus, $n_3 > NA$ and the optical admittance of this layer will be purely real. Further, we must have $|\rho_{23} \rho_{34}| = 1$. For simplicity, we consider the low-loss resist case, so that $n_2 < NA$ and the resist is in the evanescent regime (thus, its optical admittance can be approximated as purely imaginary). If the substrate is also non-absorbing and in the evanescent regime, $n_4 < NA$ and its optical admittance is also imaginary. The resulting reflectivity product is

$$\rho_{23} \rho_{34} = \left(\frac{i\eta_{2I} - \eta_{3R}}{i\eta_{2I} + \eta_{3R}}\right)\left(\frac{\eta_{3R} - i\eta_{4I}}{\eta_{3R} + i\eta_{4I}}\right) = \frac{(\eta_{2I}\eta_{4I} - \eta_{3R}^2) + i\eta_{3R}(\eta_{2I} + \eta_{4I})}{-(\eta_{2I}\eta_{4I} - \eta_{3R}^2) + i\eta_{3R}(\eta_{2I} + \eta_{4I})} \tag{22}$$

One can see that this does indeed have a magnitude of 1. Since the phase of $\tau_{D3}^2$ can be adjusted to any value by changing the thickness of layer 3, it is possible to make $\rho_{23} \rho_{34} \tau_{D3}^2 = -1$ and the overall reflectivity go to infinity. As a special case, consider $\eta_{3R} = \sqrt{\eta_{2I} \eta_{4I}}$. Putting this value into Eq. (24) gives $\rho_{23} \rho_{34} = 1$. Infinite reflectivity is obtained when $\tau_{D3}^2 = -1$, or when $D_3 = \lambda / 4\eta_{3R}$. Interestingly, these equations are identical in form to the conditions required to make a perfectly transmitting antireflection coating in the propagating regime. In the evanescent regime, they produce an infinite reflector. Our goal is not to make an infinite reflector, but one with the optimum reflectance. As discussed above, we wish for our reflectivity to be

$$\rho'_{23} \approx e^{\alpha_e D/2} \qquad (23)$$

This film stack reflectivity will have zero phase when $\tau^2_{D3} = -1$, giving

$$\rho'_{23} = \frac{\eta_{2I}\eta_{4I} + \eta^2_{3R}}{\eta_{2I}\eta_{4I} - \eta^2_{3R}} \qquad (24)$$

If $\eta_{3R}$ were adjusted to give the optimum reflectivity, the result would be

$$\eta_{3R} = \sqrt{\eta_{2I}\eta_{4I}\left(\frac{\rho'_{23}-1}{\rho'_{23}+1}\right)} \qquad (25)$$

For example, if one allowed $\alpha_e D = 4$, the optimum reflectivity would be, from Eq. (17), $\rho'_{23} \approx 7.39$. If $NA = 1.85$, $n_2 = 1.7$ and $n_4 = 1.56$ (e.g. SiO$_2$ at $\lambda = 193$ nm), the resulting optical admittance of layer 3 would be 0.743, requiring an index of refraction of $n_3 = 1.9938$ and a thickness of 65 nm, while the resist thickness would be 82.5 nm. In reality, all materials with the required optical properties may not be available. Hence we provide a real world design with some modifications. Figure 5(a) illustrates imaging of 26-nm features where the image depth is only 20 nm in an infinitely thick photoresist. Figure 5(b) illustrates the significant enhancement when a tuned stack of 43 nm layer of Al$_2$O$_3$ (sapphire) with index $n_3 = 2.08$ [11] upon ~50nm SiO$_2$ with index $n_4 = 1.56$ is used. The resulting image is 82.5 nm deep. We have also included photoresist loss in this design and the new reflection from the real world stack is ~$6.77\,e^{i0.212}$ instead of the optimum 7.39 however still allowing ultra-high $NA$ patterning at an aspect ratio of ~3.2.

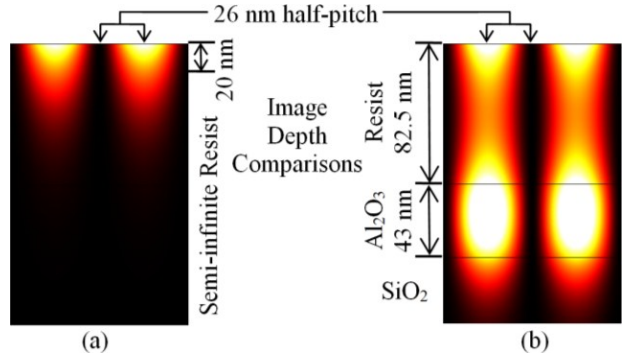

FIG. 5 Imaging of 26-nm (half-pitch) evanescent features into (a) semi-infinite lossy resist giving 20-nm image depth, and (b) 82.5 nm thick lossy resist on an effective gain medium made up of 43 nm of $Al_2O_3$ (Sapphire) on $SiO_2$, giving an image depth of 82.5 nm.

For the purposes of evanescent wave coupling at an ultra-high *NA* of 1.85, the stack behaves the same way as a hypothetical gain medium. The ATR spectrum of the system as shown in Fig. 3(b) has a resonant dip at a spatial frequency corresponding to a *NA* of ~1.83 indicative of the off-resonance operation mentioned earlier.

**CONCLUSION**

Ultra-high *NA* patterning has long been a critical outstanding problem in the semiconductor industry due to the physical limits presented by evanescent fields. We have suggested coupling of the evanescent fields to surface states as a means to control the field intensity and profile in a photoresist cavity. Through the use of the Fresnel reflection equation for TE and analysis of fields, we have presented a practical way to use an EGM surface state to image super-resolved structures with high ARs. We show how a high AR maybe achieved at *NA* = 1.85 and $\lambda$ = 193 nm in photoresist by using a layer of sapphire deposited on $SiO_2$. The method encourages use of solid-immersion systems and paves the way for feasible ultra-high-NA imaging and patterning.

We also found that any NA may be used for patterning at a tolerable aspect ratio of 1.3 by using this method. Hence, if a suitable prism material is available for evanescent interferometric imaging, it may also be deposited as a thin film on a low loss dielectric to resonate the image. We believe this finding has the potential to motivate further research in the development of higher index prism materials and immersion liquids for lithographic imaging.

**ACKNOWLEDGEMENTS**

The authors would like to acknowledge support from New Zealand's Marsden Fund (contract UOC-804), the University of Canterbury and The MacDiarmid Institute for Advanced Materials and Nanotechnology.